\documentclass[fleqn,preprint,showpacs,preprintnumbers,amsmath,amssymb]{revtex4}
\usepackage{graphicx}% Include figure files
\usepackage{dcolumn}% Align table columns on decimal point
\usepackage{bm}% bold math
\begin{document}

\title{Remarkable paramagnetic features of Fermi-Dirac plasmas}
\author{M. Akbari-Moghanjoughi}
\affiliation{Azarbaijan University of
Tarbiat Moallem, Faculty of Sciences,
Department of Physics, 51745-406, Tabriz, Iran}

\date{\today}
\begin{abstract}
In this paper by using the relativistic magnetic susceptibility of a Fermi-Dirac (relativistically degenerate) plasma, quantum magnetohydrodynamics (QMHD) model is used to investigate the propagation of spin-induced (SI) magnetosonic nonlinear excitations in a normally and relativistically degenerate dense electron-ion plasma in the presence of the spin magnetization effect. Based on the conventional pseudopotential method the matching criterion for the evolution of SI solitary structures is evaluated. It is found that, the plasma mass density and strength of the magnetic field have significant effects on excitation and evolution of magnetosonic nonlinear structures in Fermi-Dirac plasmas. Only rarefactive SI magnetosonic solitary structures are found to excite in such plasmas. Furthermore, fundamental differences are shown to exist in magnetosonic soliton dynamics in the two distinct plasma degeneracy regimes, which is due to interplay between the negative spin paramagnetism pressure-like and positive relativistic degeneracy pressure of electrons. Current investigation can help better understand the electron spin effects on nonlinear wave propagations in strongly magnetized dense astrophysical objects such as white dwarfs and pulsar magnetospheres.
\end{abstract}

\maketitle

\section{Introduction}

The subject of dense cold ionized matter under strong magnetic field has promised a wide applications in both artificial laboratory as well as naturally occurring plasmas such as astrophysical compact objects \cite{manfredi, shukla}. Such state of matter is usually termed as "quantum plasma". Quantum optics and electronic transport effects play important role in metallic and semiconductor nano-structured materials such as, nano-particles, quantum-wells, quantum-wires and quantum-dots \cite{haug}. A vast majority of recent studies on quantum plasmas \cite{bonitz, gardner, haas1, haas2, Markowich, Marklund1, Brodin1, Marklund2, Brodin2, akbari1, akbari2} have been inspired by the pioneering works of Chandrasekhar, Bohm, Pines and Levine et.al \cite{chandra1, bohm, pines, levine}. In a quantum plasma, also known as the zero-temperature Fermi-plasma the inter-fermion distances are much lower than the characteristic de Broglie thermal wavelength, $h/(2\pi m k_B T)^{1/2}$ and this causes a new type collective phenomenon due to the electron degeneracy and tunneling effect, which is encountered only in quantum plasmas such as in ordinary metals or compact astrophysical objects. The electron degeneracy is a direct consequence of the Pauli exclusion principle, which effectively rules the thermodynamical properties of a dense plasma \cite{landau}. It has been confirmed using a quantum hydrodynamics (QHD) model that, in quantum plasmas the nonlinear wave dynamics, due to delicate interplay between dissipation and dispersion, can lead to a variety of soliton-, explosive- and shock-like density structures \cite{sabry}. More recently, Marlkund and Brodin have extended QHD model to include the electron spin-1/2 effects \cite{Marklund3} by introducing a generalized term for the quantum force. Some estimates indicate that the relativistic electrons in a white dwarf can generate magnetic fields of the order of $10^7G$ and higher \cite{lee}, and that in a neutron star the self-field may be of order $10^{13}-10^{14}G$ \cite{can}. It has also been found that, in a perfect conductive quantum plasma, the spin magnetization term introduced in quantum magnetohydrodynamics (QMHD) equations acts as a negative pressure-like entity significantly effecting the dynamics of the spin-induced nonlinear magnetosonic excitations \cite{Marklund4}.

Chandrasekhar in 1935 \cite{chandra1} showed that, despite the fusion-like nature of a compact astrophysical object such as white dwarf, it can be treated a an ideal zero-temperature and completely degenerate Fermi-Dirac gas. This is due to the gigantic inward gravitational pressure which act on the star. Using the Fermi-Dirac statistics and combining it with relativity principles, in a pioneering work, he succeeded to proved that the gravitational pressure can lead to the state of a relativistic degeneracy for electrons and an ultimate collapse of the star in relativistic degeneracy case \cite{chandra2, chandra3}. The sudden collapse of the white dwarf is due to softening of the degeneracy pressure of electrons when the they become relativistic and this has been shown to alter the whole thermodynamical properties of the star \cite{kothary}. In a recent investigation it has been found that the change in the degeneracy state from nonrelativistic to relativistic one can also modify the nonlinear dynamics of a quantum plasma \cite{akbari3, akbari4}. Being inspired by the work of Marklund \cite{Marklund4} et.al, it is tempting to explore the relativistic degeneracy effects on spin-induced magnetosonic nonlinear propagations, knowing the fact that the spin pressure may have crucial effects at the level of large electron densities ($n_e\sim 10^{28}cm^{-3}$) estimated for astrophysical compact objects and also the suppressed collision rate known for Fermi-Dirac plasmas due to Pauli-blocking. The presentation of the article is as follows. The basic normalized QMHD plasma fluid equations are introduced in Sec. \ref{equations}. The nonlinear arbitrary-amplitude solution is derived in Sec. \ref{Sagdeev}. The numerical analysis is presented in Sec. \ref{discussion}. Finally, conclusions are given in section \ref{conclusion}.

\section{QMHD Model including spin magnetization}\label{equations}

We consider a collisionless quasineutral spin magnetized perfectly degenerated superdense plasma consisting of dynamic ions and fully degenerate electrons. The closed set of quantum magnetohydrodynamics equations governing the dynamics of spin-induced (SI) magnetosonic waves, taking into account the quantum tunneling and spin-1/2, in center of mass frame, contains \cite{Marklund4} the continuity equation,
\begin{equation}\label{cont}
\frac{{\partial \rho_c }}{{\partial t}} + \nabla  \cdot (\rho_c {\bf{u_c}}) = 0,
\end{equation}
the momentum equation,
\begin{equation}\label{mom}
\frac{{\partial {\bf{u_c}}}}{{\partial t}} + \left( {{\bf{u_c}} \cdot \nabla } \right){\bf{u_c}} = {\rho_c ^{ - 1}}\left( {{\bf{j}} \times {\bf{B}} - \nabla P_c + {{\bf{F}}_Q}} \right),
\end{equation}
where, the quantum force $\bm{F}_Q={\bm{F}}_B+{\bm{F}}_S$ is a collective contribution of quantum Bohm-force, ${\bm{F}}_B$, and the so-called spin-force, ${\bm{F}}_S$, namely,
\begin{equation}\label{QF}
{{\bf{F}}_Q} = \frac{{{\rho_c\hbar ^2}}}{{2{m_e}{m_i}}}\nabla \frac{{\Delta \sqrt {{\rho _c}} }}{{\sqrt {{\rho _c}} }} + {{M}} \nabla {{B}}.
\end{equation}
and the generalized Faraday law, without the Hall term,
\begin{equation}\label{FL}
\frac{{\partial {\bf{B}}}}{{\partial t}} = \nabla  \times \left\{ {{\bf{u}} \times {\bf{B}} - \eta {\bf{j}} - \frac{{{\bf{j}} \times {\bf{B}} + {{\bf{F}}_Q}}}{{e{n_e}}} - \frac{{{m_e}}}{{{e^2}{\mu _0}}}\left[ {\frac{\partial }{{\partial t}} - \left( {\frac{{\nabla  \times {\bf{B}}}}{{e{\mu _0}{n_e}}}} \right) \cdot \nabla } \right]\frac{{\nabla  \times {\bf{B}}}}{{{n_e}}}} \right\}.
\end{equation}
\textbf{The quantities $\bf{j}$, $\bf{B}$ and $P_c$ are the magnetization current, magnetic field and scalar center of mass pressure, respectively. The spin-magnetization (Pauli-magnetization) per unit volume ${{M}_P} = \left( {3\mu _B^2{n_e}/2{k_B}{T_{Fe}}} \right){{B}}$ ($T_{Fe}$ being the electron Fermi-temperature) is related to magnetization current, $\bf{j}$ through; ${\bf{j}} = \mu _0^{ - 1}\nabla  \times ({\bf{B}} - {\mu _0}{\bf{M}})$, where, $\mu_B=e\hbar/2m_e c$ is the Bohr magneton and $\hbar$ is the normalized Plank constant. Note that in the case of fully degenerate quantum plasma the Pauli paramagnetism dominates the Langevin-Type one, hence, the Fermi-temperature, $T_{Fe}$, replaces the thermal electron temperature, $T_e$ in magnetic pressure term. Due to the Pauli exclusion rule, in a completely degenerate plasma only a small fraction ($T_e/T_{Fe}$), i.e. unpaired free-electrons, with energies nearly equal to the Fermi-energy contribute to the collective spin effects. This effect gives rise to the well-known Pauli paramagnetic law of the form $M_P = 3n\mu _B^2B/(2{E_{Fe}})$ \cite{landau} instead of the Langevin-type susceptibility, $M_{Lang} = n{\mu _B}\tanh ({\mu _B}B/{k_B}{T_e})$, frequently used for dense plasmas in literature. On the other hand, the Landau demagnetization susceptibility of $M_L=-1/3M_P$ should be added due to electron orbital contribution which results in total sum of $M = M_P + M_L = n_e\mu _B^2B/{E_{Fe}}$. The susceptibility of weakly interacting relativistic Fermi-gas in a weak magnetic field ($\mu_B B_0 \ll m_e c^2$) has been evaluated in Ref. \cite{men}, which is as follows
\begin{equation}\label{QP}
{\chi _p} = \left( {\frac{{3n\mu _B^2}}{{{m_e}{c^2}}}} \right){\left( {\frac{{{m_e}c}}{{{P_{Fe}}}}} \right)^2}\sqrt {1 + {{\left( {\frac{{{P_{Fe}}}}{{{m_e}c}}} \right)}^2}} \left[ {1 + \frac{{3\alpha n}}{{2{m_e}{c^2}}}{{\left( {\frac{{{m_e}c}}{{{P_{Fe}}}}} \right)}^2}\sqrt {1 + {{\left( {\frac{{{P_{Fe}}}}{{{m_e}c}}} \right)}^2}} } \right],
\end{equation}
where, $\alpha$ and $P_{Fe}$ are the s-wave interaction parameter and the Fermi relativistic momentum, respectively. It can be shown that in the nonrelativistic limit with $\alpha=0$ the susceptibility given in Eq. (\ref{QP}) reduces to the well-known Pauli paramagnetic susceptibility.} Now, assuming the propagation of magnetoacoustic nonlinear wave to be in $x$ direction perpendicular to a magnetic field $B(x,t)$ along the $z$-axis and ignoring the Hall term in generalized Faraday law \cite{misra}, QMHD fluid equation set, assuming $\alpha=0$, may be written in the direction of propagation as \cite{Marklund4}
\begin{equation}\label{dimensional}
\begin{array}{l}
\frac{{\partial {\rho _c}}}{{\partial t}} + \frac{{\partial {\rho _c}{u_c}}}{{\partial x}} = 0,\hspace{3mm}{\rho _c} = {\rho _c}(x,t),\hspace{3mm}{u_c} = {u_c}(x,t),\hspace{3mm}B = B(x,t),\\ \frac{{\partial {u_c}}}{{\partial t}} + {u_c}\frac{{\partial {u_c}}}{{\partial x}} + \frac{B}{{{\mu _0}{\rho _c}}}\frac{{\partial B}}{{\partial x}} + \frac{1}{{{\rho _c}}}\frac{{\partial {P_c}}}{{\partial x}} - \frac{{{\hbar ^2}}}{{2{m_e}{m_i}}}\frac{\partial }{{\partial x}}\left( {\frac{1}{{{{\sqrt \rho  }_c}}}\frac{{{\partial ^2}{{\sqrt \rho  }_c}}}{{\partial {x^2}}}} \right) \\ - \frac{{3\mu _B^2}}{{{m_e}{m_i}{\rho _c}{c^2}}}\frac{\partial }{{\partial x}}\left\{ {{\rho _c}{B^2}{{\left( {\frac{{{m_e}c}}{{{P_{Fe}}}}} \right)}^2}\sqrt {1 + {{\left( {\frac{{{P_{Fe}}}}{{{m_e}c}}} \right)}^2}} } \right\} = 0, \\ \frac{{\partial B}}{{\partial t}} + \frac{{\partial B{u_c}}}{{\partial x}} - \frac{\eta }{{{\mu _0}}}\frac{{{\partial ^2}B}}{{\partial {x^2}}} = 0, \\
\end{array}
\end{equation}
where, the fourth and fifth terms in momentum equation describe the quantum force due to the Bohm potential and magnetic spin pressure gradient (spin force) due to the fermion spin-1/2 effect, respectively, with $\hbar$ being the scaled Plank constant and $\eta$ the plasma resistivity. Other parameters have their usual meanings. Assuming a quasineutrality, i.e. $n_e \simeq n_i = n$ the quantities, $\rho_c=m_e n_e + m_i n_i \simeq m_i n$, ${u_c}\simeq(m_e {u_e}+m_i {u_i})/m_i$ and $P_c$ are defined as the plasma center of mass density, speed and pressure, respectively. On the other hand, in a relativistically degenerate zero-temperature Fermi-Dirac plasma, the gigantic electron relativistic degeneracy pressure, relative to which the ion pressure is ignored, can be expressed in the following form \cite{chandra1}
\begin{equation}\label{p}
{P_e} = \frac{{\pi m_e^4{c^5}}}{{3{h^3}}}\left\{ {R\left( {2{R^2} - 3} \right)\sqrt {1 + {R^2}}  + 3\ln \left[ {R + \sqrt {1 + {R^2}} } \right]} \right\}{\rm{ }},
\end{equation}
in which $R=P_{Fe}/(m_e c)=(n/N_0)^{1/3}$ (${N_0} = \frac{{8\pi m_e^{3}{c^3}}}{{3{h^3}}}\simeq 5.9 \times 10^{29} cm^{-3}$), where $P_{Fe}$ is the electron Fermi relativistic momentum. Note that, in the limits of very small and very large values of the relativity parameter, $R$, one obtains $P_0={{{\left( {3/\pi } \right)}^{2/3}}{h^2}n_e^{5/3}/(20{m_e})}$ and $P_{\infty}={{{\left( {3/\pi } \right)}^{1/3}}hcn_e^{4/3}}/8$, respectively. Moreover, the resistivity of a completely degenerate Fermi-Dirac plasma is supposed to be negligible, since, the electron-ion collisions are suppressed by Pauli-blocking mechanism, hence, $\eta\sim 0$ which using Eq. (\ref{dimensional}) results in $B/B_0=\rho/\rho_0=n/n_0=\bar{n}$. Here, the new quantities $B_0$, $\rho_0$, $n_0$ and $\bar{n}$ denote the equilibrium values of magnetic field intensity, plasma mass-density, plasma number-density and normalized charge density, respectively. Now, using the fact that $(1/{n}){\partial _x}{P_e(n)} = {\partial _x}\sqrt {1 + R^2}$ and employing the following scalings, we may obtain the dimensionless equations
\begin{equation}\label{normal}
x \to \frac{{{c_{s}}}}{{{\omega _{pi}}}}\bar x,\hspace{3mm}t \to \frac{{\bar t}}{{{\omega _{pi}}}},\hspace{3mm}n \to \bar n{n_0},\hspace{3mm}u_c \to \bar u_c{c_{s}}.
\end{equation}
where, $c$, ${\omega _{pi}} = \sqrt {{e^2}{n_{e0}}/(\varepsilon_0{m_i})}$ and ${c_{s}} = c\sqrt {{m_e}/{(m_i)}}$ are the vacuum light speed, the characteristic ion plasma frequency and ion sound-speed (this speed is much higher despite the name comparable to the Fermi-speed of an electron in a solid), respectively, and the parameter $n_{0}$ denotes the equilibrium plasma number-density. Also the bar notation indicates the dimensionless quantities and are ignored in forthcoming algebra, for simplicity. Now, using the new compact one-dimensional quantum magnetohydrodynamics manetosonic model which includes the spin effects, we have the following two differential equations to be solved together
\begin{equation}\label{diff}
\begin{array}{l}
\frac{{\partial n}}{{\partial t}} + \frac{{\partial n{u_c}}}{{\partial x}} = 0, \\
{H^2}\frac{\partial }{{\partial x}}\left( {\frac{1}{{\sqrt n }}\frac{{{\partial ^2}\sqrt n }}{{\partial {x^2}}}} \right) = \frac{{\partial {u_c}}}{{\partial t}} + \frac{1}{2}\frac{{\partial u_c^2}}{{\partial x}} + {H^{ - 2}}{\epsilon^2}\frac{{\partial n}}{{\partial x}} + \frac{\partial }{{\partial x}}\sqrt {1 + R_0^2{n^{2/3}}}  \\
- \frac{{3{\epsilon^2}}}{2}\frac{\partial }{{\partial x}}\left[ {\ln n - 3\ln \left( {1 + \sqrt {1 + {n^{2/3}}R_0^2} } \right) - \frac{{\sqrt {1 + {n^{2/3}}R_0^2} }}{{{n^{2/3}}R_0^2}}} \right], \\
\end{array}
\end{equation}
where, we have introduced new fractional plasma entities such as the quantum diffraction parameter, $H = \sqrt {{m_i}/{2m_e}} (\hbar {\omega _{pi}})/({m_e}{c^2})$ \cite{haas2}, relativistic degeneracy parameter, $R_0=(n_0/N_0)^{1/3}$ \cite{akbari2}, and normalized Zeeman energy, $\epsilon=\mu_B B_0/(m_e c^2)$.

Before proceeding with calculations, a clear definition of the relativistic degeneracy and distinction between a low-pressure relativistic plasma from the relativistically degenerate quantum Fermi-gas is in order. The relativistic degeneracy is a completely quantum phenomenon ruled by the uncertainty principle and is increased due to the decrease in inter-fermion distances in degenerated plasmas. Although the relativistic effects arise due to increase in fermion number-density in a dense degenerate plasma, however, unlike for the low-pressure relativistic plasmas the degeneracy pressure in the fermion momentum fluid equation usually dominates the relativistic dynamic effects in super-dense plasma state. Chandrasekhar \cite{chandra2}, combining the relativity and the quantum statistics, showed that for dense degenerate Fermi-gas such as a white-dwarf with a mass-density, $\rho$, the degeneracy pressure turns from $P_e\propto \rho^{5/3}$ (with polytropic index $3$) dependence for normal degeneracy in the limit $R_0 \rightarrow 0$ to $P_e\propto \rho^{4/3}$ (with polytropic index $2/3$) dependence for relativistic degeneracy case in the limit $R_0 \rightarrow \infty$. The relativistic degeneracy starts at mass density of about $4.19\times 10^6 (gr/cm^3)$ of the order in the core a $0.3M_\odot$ white dwarf, which corresponds to a Fermi-momentum $P_{Fe}\sim 1.29 m_e c$ (equivalent to the relativistic degeneracy parameter value of $R_0\sim 1.29$) or the threshold velocity of $u_{Fe}\sim 0.63c$ (the Fermi relativistic factor $\gamma_{Fe}\sim 1.287$).

\section{Magnetosonic Nonlinear Structures}\label{Sagdeev}

In this section we seek the stationary nonlinear wave solution described by Eqs. (\ref{diff}) which is obtained using the coordinate transformation of $\xi=x-M t$, where, $M=u_c/c_s$ is the normalized matching speed of the nonlinear wave. Hence, using new variable, $n=Z^2$, Eqs. (\ref{diff}) are reduced (after integration with boundary conditions $\mathop {\lim }\limits_{\xi  \to  \pm \infty } n = 1$ and $\mathop {\lim }\limits_{\xi  \to  \pm \infty } u_c = 0$) to the following differential equation
\begin{equation}\label{diff2}
\begin{array}{l}
\frac{{{H^2}}}{Z}\frac{{{\partial ^2}Z}}{{\partial {\xi^2}}} = \frac{{{M^2}}}{2}{\left( {1 - {Z^{ - 2}}} \right)^2} - {M^2}\left( {1 - {Z^{ - 2}}} \right) + \frac{\epsilon^2}{H^{ 2}}\left( {{Z^2} - 1} \right) + \sqrt {1 + R_0^2{Z^{4/3}}}  - \sqrt {1 + R_0^2}  \\ - \frac{{3{\epsilon^2}}}{2}\left[ {\ln n - 3\ln \left( {1 + \sqrt {1 + {n^{2/3}}R_0^2} } \right) - \frac{{\sqrt {1 + {n^{2/3}}R_0^2} }}{{{n^{2/3}}R_0^2}} + 3\ln \left( {1 + \sqrt {1 + R_0^2} } \right) - \frac{{\sqrt {1 + R_0^2} }}{{R_0^2}}} \right]. \\
\end{array}
\end{equation}
Now multiplying both sides of Eq. (\ref{diff2}) with $dZ/d\xi$ and integrating with aforementioned boundary conditions, we get the well known energy integral of the form
\begin{equation}\label{energy}
{({d_\xi }n)^2}/2 + U(n) = 0,
\end{equation}
with the pseudopotential given as
\begin{equation}\label{pseudo}
\begin{array}{l}
U(n) = \frac{n}{{4{H^2}}}\left[ {\frac{{4{M^2}}}{n} - 8{M^2} + 4{M^2}n - 2\sqrt {1 + {R_0^2}}  + 8n\sqrt {1 + {R_0^2}}  + \frac{{3\sqrt {1 + {R_0^2}} }}{{{R_0^2}}}} \right. \\
- 6n\sqrt {1 + {n^{2/3}}{R_0^2}}  - \frac{{3{n^{1/3}}\sqrt {1 + {n^{2/3}}{R_0^2}} }}{{{R_0^2}}} - \frac{{4{\epsilon^2}}}{{{H^2}}} + \frac{{8{\epsilon^2}n}}{{{H^2}}} - \frac{{4{\epsilon^2}{n^2}}}{{{H^2}}} - \frac{{12{\epsilon^2}\sqrt {1 + {R_0^2}} }}{{{R_0^2}}} \\ + 36{\epsilon^2}n\ln (1 + \sqrt {1 + {R_0^2}} ) + 36{\epsilon^2}\ln (1 + \sqrt {1 + {n^{2/3}}{R_0^2}} ) - 36{\epsilon^2}n\ln (1 + \sqrt {1 + {n^{2/3}}{R_0^2}} ) \\ + \frac{{3{{\sinh }^{ - 1}}({n^{1/3}}R_0)}}{{{R_0^3}}} - 12{\epsilon^2}\ln n + 12{\epsilon^2}n\ln n - 36{\epsilon^2}\ln (1 + \sqrt {1 + {R_0^2}} ) \\ \left. { + \frac{{12{{\epsilon^2}}n\sqrt {1 + {R_0^2}} }}{{{R_0^2}}} + \frac{{12{{\epsilon^2}}\sqrt {1 + {n^{2/3}}{R_0^2}} }}{{{n^{2/3}}{R_0^2}}} - \frac{{12{\epsilon^2}{n^{1/3}}\sqrt {1 + {n^{2/3}}{R_0^2}} }}{{{R_0^2}}} - \frac{{3{{\sinh }^{ - 1}}R_0}}{{{R_0^3}}}} \right] \\
\end{array}
\end{equation}
Note that the quantum diffraction parameter, $H$, and the relativistic degeneracy parameter, $R_0$, are not independent parameters but are interrelated through $H = e\hbar \sqrt {{N_0}R_0^3/\pi } /(2m_e^{3/2}{c^2})$. The possibility of solitary excitation relies on some conditions to satisfy, simultaneously, namely
\begin{equation}\label{conditions}
{\left. {U(n)} \right|_{n = 1}} = {\left. {\frac{{dU(n)}}{{dn}}} \right|_{n = 1}} = 0,\hspace{3mm}{\left. {\frac{{{d^2}U(n)}}{{d{n^2}}}} \right|_{n = 1}} < 0.
\end{equation}
It is further required that for at least one either maximum or minimum nonzero $n$-value, we have $U(n_{m})=0$, so that for every value of $n$ in the range ${n _m} > n  > 1$ (compressive soliton) or ${n _m} < n  < 1$ (rarefactive soliton), $U(n)$ is negative (it is understood that there is no root in the range $[1,n_m]$). In such a condition we can obtain a potential minimum which describes the possibility of a solitary wave propagation. The stationary soliton solutions corresponding to this pseudo-potential which satisfies the mentioned boundary-conditions, read as
\begin{equation}\label{soliton}
\xi  - {\xi _0} =  \pm \int_1^{n_m} {\frac{{dn}}{{\sqrt { - 2U(n)} }}}.
\end{equation}
The conditions for the existence of a solitary propagation stated above require that, first, it takes infinitely long pseudo-time ($\xi$) for the system to get away from the unstable point ($n=1$). This statement requires that $d_n U(n)\mid_{n=1}=0$ or equivalently $d_\xi n\mid_{\xi=-\infty}=0$ in parametric space, as it is also inferred by the shape of a solitary wave. Thereafter, moving forward in pseudo-time ($\xi$) axis, the localized density perturbation reaches a maximum or a minimum at $n=n_m$ (if it exists) at which the pseudo-speed ($d_\xi n$) of the analogous particle bound in pseudo-potential ($U(n)$) region of $1>n>n_m$ (or $1<n<n_m$) reaches zero again and it returns back. Note that, in the parametric space, from equation Eq. (\ref{soliton}), it is observed that in physical situation $U(n)$ should be negative for solitary (non-periodic) wave solution, which is clearly satisfied if $d_{nn}U(n)\mid_{n=1}<0$ and $U(n_m\neq 1)=0$. Note also that, both the requirements $U(n)\mid_{n=1}=0$ and $d_{n}U(n)\mid_{n=1}=0$ follow from the equilibrium state assumption at infinite pseudo-time ($\xi=\pm\infty$) before and after perturbation takes place, i.e. $d_{\xi\xi} n\mid_{\xi=\pm\infty}=d_{\xi} n\mid_{\xi=\pm\infty}=0$. However, there is a special case with $d_{n}U(n)\mid_{n=n_m}=0$ for which the density perturbation is stabilized at the maximum or minimum density $n=n_m$ (the analogous particle never returns back). This situation regards to the existence of a double-layer in plasma which is not considered here. It can be confirmed that, the pseudopotential given by Eq. (\ref{pseudo}) and its first derivative vanish at $n=1$, as required by the two first conditions, Eq. (\ref{conditions}). Also, direct evaluation of the second derivative of the Sagdeev potential, Eq. (\ref{pseudo}), at unstable point, $n=1$, leads to
\begin{equation}\label{dd}
{\left. {\frac{{{d^2}U(n)}}{{d{n^2}}}} \right|_{n = 1}} = \frac{{{H^2}\left[ {12{\epsilon^2} - 2{R_0^4} + 6{R_0^2}\left( {{M^2}\sqrt {1 + {R_0^2}}  + 4{\epsilon^2}} \right)} \right] - 6{\epsilon^2}{R_0^2}\sqrt {1 + {R_0^2}} }}{{3{H^4}{R_0^2}\sqrt {1 + {R_0^2}} }}.
\end{equation}
that is, for the existence of a SI magnetosonic solitary excitation the soliton matching number should be below a threshold Mach-value defined below
\begin{equation}\label{consol2}
M_{cr} = \frac{{\sqrt {3{\epsilon^2}R_0^2\sqrt {1 + R_0^2}  + {H^2}\left[ {R_0^4 - 6{\epsilon^2}(1 + 2R_0^2)} \right]} }}{{\sqrt 3 H{R_0}{{(1 + R_0^2)}^{1/4}}}}.
\end{equation}
It is further required that for at least one either maximum or minimum nonzero $n_m$-value, we have $U(n_m)=0$, so that for every value of $n$ in the range ${n_m} > n  > 0$ or ${n_m} < n  < 0$, $U(n)$ is negative. In such a condition there will be a potential minimum which describes the propagation of solitary nonlinear structure. To this end, we evaluate the existence of $n_m$ values which is essential to our analysis. A close inspection of the pseudopotential given in Eq. (\ref{pseudo}) reveals that
\begin{equation}\label{nm}
\mathop {\lim }\limits_{n \to 0} U(n) = \frac{{{M^2}}}{{{H^2}}} > 0,\hspace{3mm}\mathop {\lim }\limits_{n \to \infty } U(n) =  - \infty,
\end{equation}
a result which is independent of all plasma parameters. This leads to the conclusion that only rarefactive ($n_m<1$) soliton may exist in our model a situation which is similar to the results presented in Ref. \cite{Marklund4, akbari3}. Note also that Eq. (\ref{consol2}) agrees well with the findings of Ref. \cite{akbari6} in the limit $\epsilon=0$. Also, Eq. (\ref{nm}) is in agreement with the findings of Ref. \cite{akbari7}. However, in the model considered in Ref. \cite{akbari6} without the spin pressure only compressive ion acoustic solitary structures has been found.

\section{Numerical Analysis and Discussion}\label{discussion}

\textbf{In Fig. 1, we shows the paramagnetic susceptibility $\chi_p$, given in Eq. (\ref{QP}), versus the relativity parameter, $R=(n_e/N_0)^{1/3}$ for a relativistically degenerate Fermi-gas (Fermi-Dirac plasma). It is clearly observed that the increase in the relativity parameter (relativistic degeneracy of the plasma) causes the increase in the relativistic susceptibility value. This is due to the increase in the number density of electrons.}

\textbf{The soliton stability region is shown in Fig. 2, which indicates that the existence of magnetoacoustic nonlinear propagation is significantly affected by the relativistic degeneracy and the Pauli spin magnetization effects. In this plot and others mentioned further we have used appropriate range for the Zeeman energy parameter, $10^{-7}<\epsilon<10^{-5}$ which corresponds to the field strength range, $10^3<B_0(T)<10^5$, typical of astrophysical compact objects. However, for some neutron stars or pulsars the corresponding magnetic field may even be much higher ($B_0\simeq 10^8T$). The critical Mach curve of $M$-$R_0$ plane at $\epsilon=0$ or $B_0=0$ is clearly consistent with the plots given in Ref. \cite{akbari6} and indicates the increase in the soliton matching speed-range with increase in the value of relativistic degeneracy parameter, $R_0$. On the other hand, the introduction of a magnetic field ($\epsilon\neq 0$) change the lower part of $M$-$R_0$ cross-section (nonrelativistic plasma degeneracy regime, i.e. $R_0\ll 1$) effectively, leaving the large $R_0$ (relativistic plasma degeneracy regime, i.e. $R_0> 1$) part almost unaffected. This means that the soliton speed for denser plasmas are less affected by Pauli paramagnetic effects. Furthermore, for every nonzero value of $\epsilon$ there is minimum soliton matching speed, $M_m$, in $M$-$R_0$ plane, value of which increases as the strebght of the magnetic field increases. As it is observed from Fig. 2, the Pauli spin paramagnetism causes the soliton matching speed-range to increase to higher values. It is remarkable that, as the paramagnetic effect increases (magnetic field becomes stronger) the widening of the soliton matching speed-range increases and extends to higher plasma mass-density area.
}
\textbf{Figure 3 points to remarkable differences of pseudopotential for soliton existence in different plasma degeneracy regimes, namely, nonrelativistic and relativistic degeneracies. First of all is the difference in the scale of the pseudopotential itself which is related to the soliton width. The second is the effect of the relativistic degeneracy parameter, $R_0$ on the potential width, while the effect is similar for the potential depth (compare Figs. 3(a) and 3(b)). Third difference is noticed easily comparing Figs. 3(c) and 3(d) where shows the ineffectiveness of the magnetic field change on the potential profile in the relativistic degeneracy case regarding to that of normal degeneracy. However, there is almost no difference in the effect of change in matching soliton speed on potentials between the two mentioned cases, since, this is related to the general properties of nonlinear wave propagation theory in dispersive media.
}
\textbf{In Fig. 4 we have shown the effect of variation of plasma fractional parameter on the width and amplitude of the magnetoacoustic solitons in a quantum plasma in the presence of Pauli spin magnetization. In this figure the differences between the two extreme degeneracy regimes are obvious. It is noted that the soliton widths are comparably larger in relativistically degenerate plasma (left column) compared to those of the normally degenerate ones (right column). As it is remarked by comparing Figs. 4(a) and 4(b) the increase in the plasma mass-density, $\rho_0$ (or equivalently the value of relativistic degeneracy parameter, $R_0$) increases/decreases the soliton amplitude in relativistic/nonrelativistic plasma degeneracy regime while the soliton width always increases with increase in the plasma mass-density value for both relativistic and nonrelativistic degeneracy regimes. It appears from Figs. 4(b) that the magnetic field has absolutely no effect on soliton properties in a relativistically degenerate plasma compared to its partner in norelativistic one, shown in Fig. 4(d), which its amplitude/wicth increases/decreases due to increase in the local magnetic field strength, $B_0$ (or increase in normalized Zeeman energy). This property marks a distinct feature of soliton dynamics in superdense astrophysical objects which requires further investigations. Finally, it is confirmed by Figs. 4(e) and 4(f) that, the faster solitons are smaller and wider in both plasma degeneracy regimes, consistent with the soliton propagation properties.}

\section{Conclusions}\label{conclusion}

We investigated the effect of Pauli spin magnetization on nonlinear SI magnetosonic wave dynamics in both relativistic and nonrelativistically degenerate quantum plasma. The standard pseudopotential method is employed to achieve the matching condition for propagation of these waves based on quantum magnetohydrodynamics equations. It was confirmed that, independent of the values of plasma fractional parameters such as wave Mach-number, relativistic degeneracy parameter and the strength of the perpendicular magnetic field, always rarefactive solitary structures exist. It was revealed that fundamental differences in soliton dynamics propagating in the two plasma degeneracy regimes exist. Current results can be useful in the study of astrophysical magnetized compact objects such as white dwarfs, pulsar magnetospheres and interiors of giant planets.

\newpage

\textbf{FIGURE CAPTIONS}

\bigskip

Figure-1

\bigskip

The paramagnetic susceptibility, $\chi_p$, versus the relativity parameter, $R=(n_e/N_0)^{1/3}$, for a relativistically degenerate Fermi-gas (Fermi-Dirac plasma).

\bigskip

Figure-2

\bigskip

The volume confined to thick borders shows the region where the solitary magnetosonic structures can exist. The principle axis are the matching soliton speed, $M$, the relativistic degeneracy parameter, $R_0$ related to solely to the plasma mass-density, $\rho_0$, and the normalized Zeeman energy, $\epsilon$ which depends only to the local magnetic field strength, $B_0$.

\bigskip

Figure-3

\bigskip

(Color online) Figure (2) depicts the profiles and variations of pseudopotential depth and width for the rarefactive solitary spin-induced magnetosonic solitary waves with respect to change in each of three independent plasma fractional parameter, namely, normalized wave speed, $M$, normalized relativity parameter, $R_0$, and normalized Zeeman energy $\epsilon$, while the other two parameters are fixed. The left column presents the pseudopotential for the relativistic degeneracy plasma regime and the right column belongs to the nonrelativistic plasma degeneracy regime. The axis of pseudopotential $U$ in the right column is normalized to a appropriate factor shown in each plot. The dash size of pseudopotential increases appropriately as the varied parameter in each plot is increased.

\bigskip

Figure-4

\bigskip

(Color online) Figure 3 shows the soliton profiles and their variation with respect to change in each of three independent plasma fractional parameter, namely, normalized wave speed, $M$, normalized relativity parameter, $R_0$, and normalized Zeeman energy $\epsilon$, while the other two parameters are fixed. The left column presents the pseudopotential for the relativistic degeneracy plasma regime and the right column belongs to the nonrelativistic plasma degeneracy regime. The thickness of profiles increases appropriately as the varied parameter in each plot is increased.

\bigskip

\newpage

\begin{figure}[ptb]\label{Figure1}
\includegraphics[scale=.7]{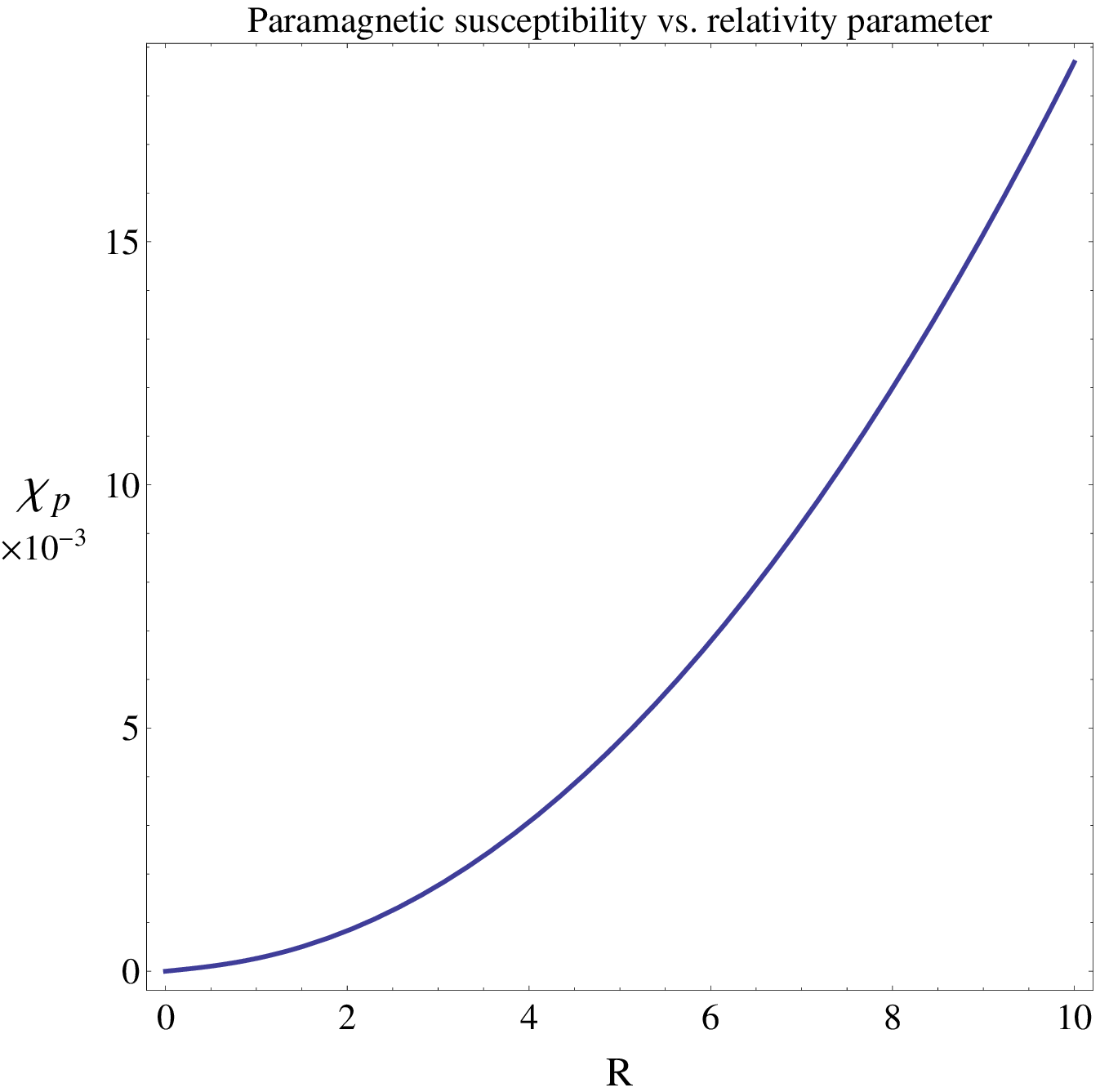}\caption{}
\end{figure}

\newpage

\begin{figure}[ptb]\label{Figure1}
\includegraphics[scale=.7]{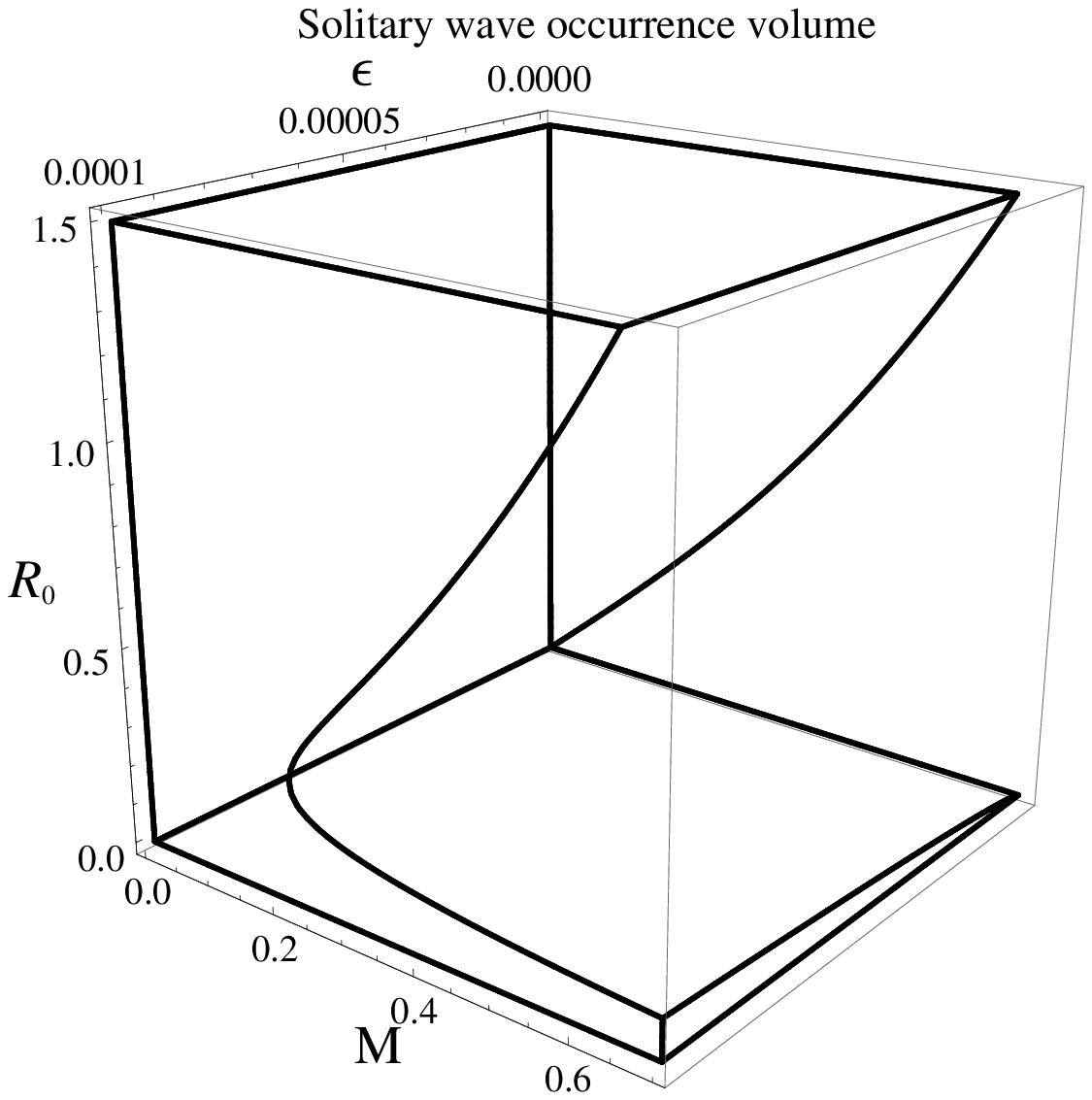}\caption{}
\end{figure}

\newpage

\begin{figure}[ptb]\label{Figure2}
\includegraphics[scale=.5]{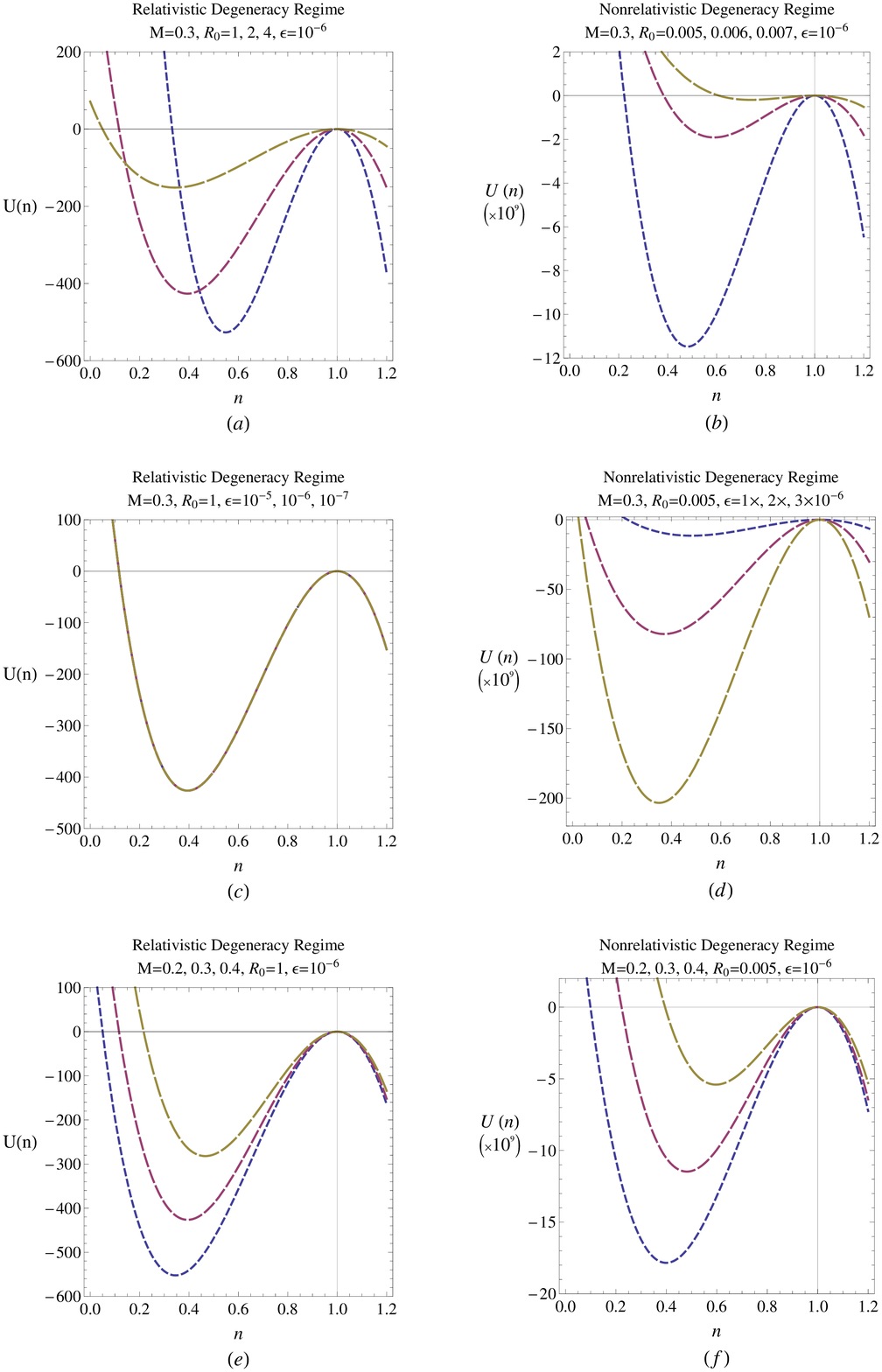}\caption{}
\end{figure}

\newpage

\begin{figure}[ptb]\label{Figure3}
\includegraphics[scale=.5]{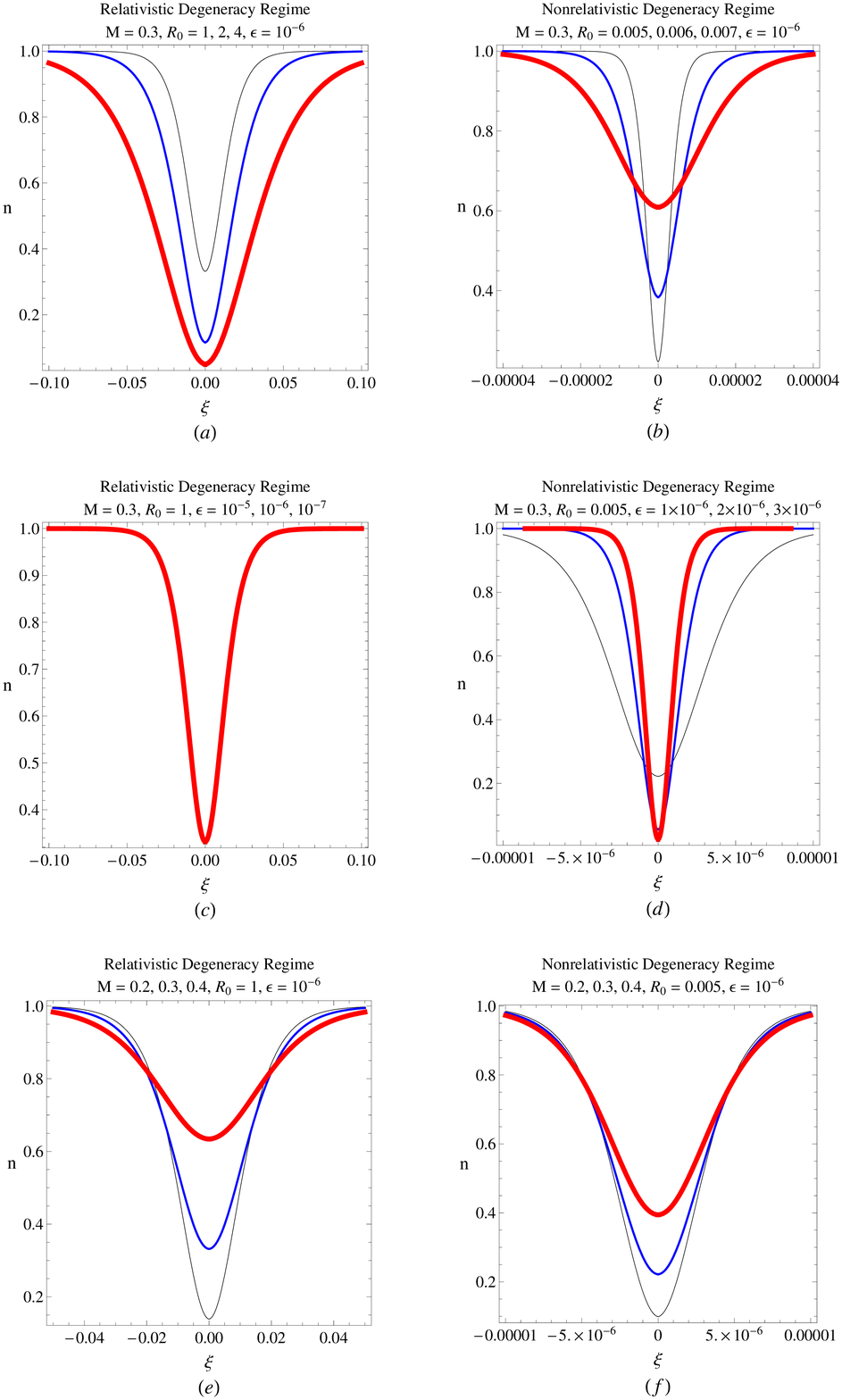}\caption{}
\end{figure}

\end{document}